# Flow Model of Intersystem Interactions and Influence of Components of Multilayer Network Systems

Olexandr Polishchuk

Laboratory of Modeling and Optimization of Complex Systems
Pidstryhach Institute for Applied Problems of Mechanics and Mathematics, National Academy of Sciences of Ukraine, Lviv, Ukraine
od_polishchuk@ukr.net

**Abstract** – *The approach to decomposition of multidimensional network systems on monoflow multilayer network systems (MLNS) is proposed, which significantly simplifies the analysis of inter-system interactions in multipurpose and multifunctional supersystem formations. A flow model of monoflow partially overlapped MLNS is built, which reflects the process of intersystem interactions between network systems that ensure the movement of flows of a certain type. The main local and global flow characteristics of MLNS elements are determined, which allow us more adequately from a functional point of view to establish the importance of nodes and edges of a multilayer system compared to their structural characteristics. The constructed model is proposed to be used for the development of methods of protecting complex network and multilayer network systems from targeted attacks and non-targeted damages of various types and solving other practically important problems of system analysis.*

**ВСТУП**

У статті [1] були проаналізовані основні види негативних внутрішніх та зовнішніх впливів на складні мережеві системи (МС) та процеси міжсистемних взаємодій. Серед таких впливів насамперед були виділені цілеспрямовані атаки та нецільові ураження складних систем, які можуть мати локальний, груповий або загальносистемний характер та бути спрямованими на ураження як структури, так і процесу функціонування мережевих та багатошарових мережевих систем (БШМС) різних типів. У статті [2] на підставі структурної моделі БШМС були запропоновані сценарії послідовних та одночасних групових та загальносистемних атак на структуру міжсистемних взаємодій. Однак, структурні показники далеко не завжди адекватно відображають функціональну важливість окремих складових БШМС, а отже і структурні сценарії, які базуються на таких показниках, можуть виявитися не самими ефективними. Безумовно, ураження структури тим або іншим чином дестабілізує процес функціонування системи, але це може відбуватися і за неураженої структури. Так, знищення російським агресором нафтобаз в Україні у травні-червні 2022 р. призвело до значних труднощів у постачанні пального, а електропідстанцій у жовтні-грудні 2022 р. – до тривалих перебоїв у постачанні електроенергії, води та тепла, зникнення мобільного зв'язку та Інтернету тощо. Водночас, блокування морських портів України навесні та влітку 2022 р. під час російсько-української війни без їх безпосереднього ураження суттєво зменшило обсяги міжнародних вантажних перевезень продукції українських аграріїв та металургів, а припинення роботи аеропортів із-за небезпеки авіаперельотів – до перерозподілу пасажирських потоків автомобільним та залізничним транспортом. У статті [3] було показано, що на підставі потокової моделі МС можна значно адекватніше визначати важливість елементів у процесі функціонування відповідної системи і, унаслідок цього, будувати значно дієвіші сценарії цілеспрямованих атак на неї. Корисність таких сценаріїв полягає у тому, що вони, даючи картину можливого розвитку атаки, дозволяють розробляти ефективні засоби захисту від неї. До функціональних показників важливості елемента МС насамперед належать параметри впливу та посередництва вузлів та ребер мережевої системи [4], які дають змогу не лише визначати складові системи, ураження яких завдадуть їй найбільшої шкоди, але й кількісно оцінити рівень цієї шкоди.

Процеси, які перебігають під час міжсистемних взаємодій, є значно глибшими та складнішими [5]. Успішна цілеспрямована атака на одну із МС, яка приймає участь у таких взаємодіях, або її нецільове ураження тим або іншим чином негативно впливає на процес функціонування усіх пов'язаних із нею систем, як це сталося під час поширення пандемії Covid-19 [6]. Водночас, існують загальносистемні атаки на окремі або всі шари БШМС, до яких можна віднести гібридні війни або комплексні економічні, фінансові, політичні та інші санкції, які застосовуються до країн, що несуть загрозу світовій безпеці [7]. Світова спільнота виявилася неготовою до своєчасної протидії та подоланню наслідків таких загальносистемних уражень [8]. Це свідчить про необхідність розроблення методів, які допомагали б глибше розуміти подібні процеси та захищати структуру та процес функціонування як окремих мережевих систем, так і міжсистемних взаємодій різних типів.

**Мета статті** – розроблення потокової моделі міжсистемних взаємодій та визначення на її підставі функціональних показників важливості елементів БШМС і формування ефективних сценаріїв послідовних та одночасних групових та загальносистемних цілеспрямованих атак на процес функціонування монопотокових частково покритих багатошарових мережевих систем.

## ПОТОКОВА МОДЕЛЬ БАГАТОШАРОВОЇ МЕРЕЖЕВОЇ СИСТЕМИ

Структурна модель міжсистемних взаємодій описується багатошаровими мережами (БШМ) та відображається у вигляді [2, 9]

$$G^M = \left( \bigcup_{m=1}^{M} G_m, \bigcup_{\substack{m,k=1 \\ m \neq k}}^{M} E_{mk} \right), \qquad (1)$$

де $G_m = (V_m, E_m)$ визначає структуру $m$-го мережевого шару БШМ; $V_m$ – множина вузлів мережі $G_m$; $E_m$ – множина зв'язків мережі $G_m$, $E_{mk}$ – множина зв'язків між вузлами множин $V_m$ та $V_k$, $m \neq k$, $m,k = \overline{1,M}$, $M$ – кількість шарів БШМ. Множину

$$V^M = \bigcup_{m=1}^{M} V_m$$

називатимемо загальною сукупністю вузлів БШМ, $N^M$ – кількість елементів $V^M$, а множини

$$E_{intra}^M = \bigcup_{m=1}^{M} E_m, \quad E_{inter}^M = \bigcup_{\substack{m,k=1 \\ m \neq k}}^{M} E_{mk}$$

загальними сукупностями внутрішньо та міжшарових зв'язків БШМС (із врахуванням орієнтації) відповідно, $L_{intra}^M$ та $L_{inter}^M$ – кількості елементів множин $E_{intra}^M$ та $E_{inter}^M$. Позначимо

$$E^M = E_{intra}^M \bigcup E_{inter}^M,$$

$L^M = L_{intra}^M + L_{inter}^M$ – загальна кількість зв'язків багатошарової мережі $G^M$.

Багатошарова мережа $G^M$ повністю описується матрицею суміжності $\mathbf{A}^M = \{\mathbf{A}^{km}\}_{m,k=1}^M$, у якій значення $a_{ij}^{km} = 1$, якщо існує ребро, яке з'єднує вузли $n_i^k$ та $n_j^m$, та $a_{ij}^{km} = 0$, $i,j = \overline{1, N^M}$, якщо такого ребра немає. Блоки $\mathbf{A}^{km} = \{a_{ij}^{km}\}_{i,j=1}^{N^M}$, $m,k = \overline{1,M}$, матриці $\mathbf{A}^M$ визначаються для загальної сукупності вузлів БШМ, тобто знімається проблема координації номерів вузлів у випадку їх незалежної нумерації для кожного шару. Зі структурного погляду найбільш загальним видом багатошарових мереж можна вважати частково покриті (ЧП) БШМ, перетин множин вузлів $V_m$ яких є непорожнім (рис. 1) [10].

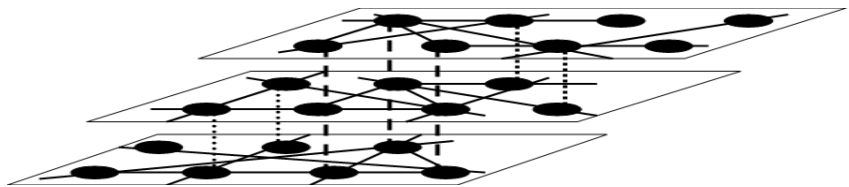

Рис. 1. Приклад структури частково покритої багатошарової мережі

Більшість реально існуючих систем та міжсистемних взаємодій є багатоцільовими та багатофункціональними. Це насамперед виражається у мультипотоковості таких утворень, тобто забезпеченні руху різних типів потоків. У теорії складних мереж (ТСМ) структура по-дібних міжсистемних взаємодій відображається так званими багатовимірними мережами [11]. Структура багатовимірної мережі має вигляд БШМ, у якій кожний шар відображає структуру системи, що забезпечує рух заданого типу потоку. Прикладами багатовимірних мереж є структура міжнародного співробітництва, до складу якої входять шари політичного, фінансового, економічного, військового, безпекового, культурного, спортивного співробітництва і т. ін., структура системи забезпечення життєдіяльності великого міста, яка включає шари електро-, водо-, тепло-, газопостачання, надання послуг телебачення, мобільного та стаціонарного телефонного зв'язку, Інтернету тощо. Розглянемо у якості прикладу багатовимірної мережевої системи загальну транспортну систему (ЗТС), яка забезпечує рух двох основних видів потоків – пасажирських та вантажних, тобто її структуру можна зобразити у вигляді двовимірної мережі. Особливістю цієї структури, як і більшості багатовимірних мереж, є неможливість переходу потоку з одного шару на інший (перетворення пасажирів у вантажі і навпаки). Для спрощення подальшого аналізу процесу міжсистемних взаємодій у двовимірній ЗТС її можна поділити на чотири двовимірні мережі, кожна з яких забезпечує перевезення залізничним, автомобільним, авіаційним та водним (морським і річковим) видами транспорту відповідно. Незважаючи на те, що перехід потоків між шарами таких мереж є неможливим, ці шари є взаємодіючими, оскільки, наприклад, для залізниці необхідно узгоджувати графіки руху пасажирських та вантажних поїздів, використання локомотивів, системи зв'язку, роботу персоналу станцій та систем управління пасажирськими та вантажними перевезеннями і т. ін. Інший підхід полягає у зображенні кожного із шарів двовимірної ЗТС у вигляді чотиришарової монопотокової БШМС, шари якої (залізничний, автомобільний, авіаційний та водний) забезпечує рух лише одного типу потоку – пасажирського або вантажного. Характерною рисою монопотокових (МП) транспортних БШМС є відмінність носіїв потоків у кожному шарі (поїзди, автотранспортні засоби, літаки, кораблі). Можливість використання різних носіїв потоків не тільки сприяє потенційному збільшенню швидкості їхнього руху, здешевленню перевезень або можливості доступу до вузлів мережі, недосяжних в окремих її шарах, але й виконує функцію дублювання шляхів доступу між найважливішими вузлами ЗТС, що підвищує надійність процесу її функціонування. Загалом під час деталізації структури реальних багатовимірних мереж спочатку доцільно виділяти шари, які забезпечують рух різних типів потоків, а потім кожний із таких монопотокових шарів зображати у вигляді БШМС кожний шар якої забезпечує рух цих потоків специфічним носієм. У цій статті ми розглядаємо випадок, коли міжшарові зв'язки у монопотокових БШМС можуть існувати лише між вузлами з однаковими номерами загальної сукупності вузлів, тобто кожний вузол може бути елементом кількох систем та виконувати у них одну функцію, але різними способа-ми (рис. 1). Вузли, у яких можливий перехід потоку з одного шару на інший, називатимемо точками переходу БШМС.

Загалом під потоком, який проходить ребром мережі, ми розуміємо певну, співвіднесену до цього ребра, додатну функцію. Ця функція може відображати щільність потоку у кожній точці ребра або об'єм потоку, який знаходиться на ребрі у поточний момент часу $t \geq 0$, чи сумарний об'єм потоку, який пройшов ребром мережі до поточного моменту за певний період тривалістю $T>0$ і т. ін. Відобразимо сукупність потоків, які проходять усіма ребрами монопотокової БШМС, у вигляді потокової матриці суміжності $\mathbf{V}^M(t)$, елементи якої визначаються об'ємами потоків, які пройшли ребрами БШМ (1) за період $[t-T, t]$ до поточного моменту часу $t \geq T$:

$$\mathbf{V}^M(t) = \{V_{ij}^{km}(t)\}_{i,j=1}^N, \underset{k,m=1}{\overset{M}{}} \quad V_{ij}^{km}(t) = \frac{\widetilde{V}_{ij}^{km}(t)}{\max\limits_{s,g=\overline{1,M}} \max\limits_{l,p=\overline{1,N^M}} \{\widetilde{V}_{lp}^{sg}(t)\}}, \quad (2)$$

де

$$\widetilde{V}_{ij}^{km}(t) = \int_{t-T}^{t} v_{ij}^{km}(\tau)d\tau, \quad t \geq T > 0; \quad v_{ij}^{km}(t) = \int_{(n_i^k, n_j^m)} \rho_{ij}^{km}(t, \mathbf{x})dl; \quad \boldsymbol{\rho}(t,x) = \{\rho_{ij}^{km}(t, \mathbf{x})\}_{i,j=1}^{N^M}, \underset{k,m=1}{\overset{M}{}}$$

і $\rho_{ij}^{km}(t, \mathbf{x})$ – щільність потоку, який пересувається ребром $(n_i^k, n_j^m)$ БШМС у поточний момент часу $t$, $\mathbf{x} \in (n_i^k, n_j^m) \subset R^n$, $n = 2, 3, \ldots$, $i, j = \overline{1, N^M}$, $k, m = \overline{1, M}$, $t > 0$. Очевидно, що структура матриці $\mathbf{V}^M(t)$ співпадає зі структурою матриці $\mathbf{A}^M$. Елементи потокової матриці суміжності БШМС визначаються на підставі емпіричних даних про рух потоків її ребрами. Натепер за допомогою сучасних засобів відбору інформації такі дані достатньо легко отримати для багатьох природних та переважної більшості створених людиною систем (транспортних, енергетичних, фінансових, інформаційних тощо) [12]. Матриця $\mathbf{V}^M(t)$ аналогічно $\mathbf{A}^M$ також має блочну структуру, у якій діагональні блоки $\mathbf{V}^{mm}(t)$ описують об'єми руху внутрішньо шарових потоків у $m$-му шарі, а позадіагональні блоки $\mathbf{V}^{km}(t)$ – об'єми руху потоків між $m$-тим та $k$-тим шарами БШМС, $m \neq k$, $m,k = \overline{1,M}$. Зрозуміло, що описана вище потокова модель БШМС не є її математичною моделлю у звичному розумінні цього слова, але вона дає достатньо чітке кількісне уявлення про процеси внутрішньо та міжсистемних взаємодій, дозволяє аналізувати особливості та прогнозувати поведінку цих процесів, а також оцінювати їх ефективність та запобігати існуючим або потенційним загрозам [13, 14]. Такий підхід обраний тому що побудова математичних моделей руху потоків реальною складною мережею наштовхується на суттєві труднощі. Наприклад, відобразити рух потоків за допомогою гідродинамічної моделі з вузла-генератора через ребро (трубу) у вузол-приймач у вигляді звичайно-го диференціального рівняння достатньо просто [15]. Однак, якщо мережа містить десятки або сотні тисяч вузлів, кожний з яких може одночасно бути і генератором, і транзитером, і приймачем потоків, об'єми яких неперервно змінюються, та десятки або сотні тисяч ребер (що і визначає розмірність системи диференціальних рівнянь гідродинамічної моделі) є достатньо складною проблемою навіть для сучасних комп'ютерних технологій [16, 17].

**ПОТОКОВІ ХАРАКТЕРИСТИКИ ЕЛЕМЕНТІВ БАГАТОШАРОВИХ МЕРЕЖЕВИХ СИСТЕМ**

Визначимо найважливіші локальні потокові характеристики елементів МП ЧП БШМС. Під локальною розумітимемо характеристику, яка описує властивості самого елемента або той чи інший аспект його взаємодії з безпосередньо пов'язаними (суміжними) елементами системи. Локальною потоковою характеристикою ребра $(n_i^k, n_j^m)$ вважатимемо відповідний елемент потокової матриці суміжності $\mathbf{V}^M(t)$, тобто об'єм потоків, які пройшли через це ребро за проміжок часу $[t-T,t], t \geq T$. Зазвичай локальною або глобальною характеристикою вузла багатошарової мережі у ТСМ вважається вектор його локальних або глобальних характеристик в окремих шарах БШМ [9]. Оскільки об'єми потоків, які надходять у певний вузол багатошарової системи із суміжних вузлів або спрямовуються із цього вузла до них можуть суттєво відрізнятися, то вхідний та вихідний потокові ступені $\boldsymbol{\delta}_i^{in,intra}(t)$ та $\boldsymbol{\delta}_i^{out,intra}(t)$ вузла $n_i$ загальної сукупності вузлів $V^M$ багатошарової мережевої системи визначаються, як

$$\boldsymbol{\delta}_i^{in,intra}(t) = \{\delta_{m,i}^{in,intra}(t)\}_{m=1}^M, \quad \boldsymbol{\delta}_i^{out,intra}(t) = \{\delta_{m,i}^{out,intra}(t)\}_{m=1}^M, \qquad (3)$$

у яких значення

$$\delta_{m,i}^{in,intra}(t) = \sum_{j=1}^{N^M} V_{ji}^{mm}(t), \quad \delta_{m,i}^{out,intra}(t) = \sum_{j=1}^{N^M} V_{ij}^{mm}(t) \qquad (4)$$

визначають вхідні та вихідні потокові ступені вузла $n_i$ загальної сукупності вузлів у $m$-му шарі відповідно, $i = \overline{1, N^M}$. У формулах (4) враховується, що зв'язки-петлі у шарах БШМС виключаються, тобто діагональні елементи матриць $\mathbf{V}^{mm}(t)$ є нульовими, $m = \overline{1, M}$, $t \geq T$. Такий підхід є цілком виправданим під час дослідження багатовимірних МС [12], оскільки об'єми потоків, які проходять через даний вузол у певному шарі таких багатошарових систем загалом не залежать від об'ємів потоків, які проходять через нього у інших її шарах. Однак, під час дослідження властивостей монопотокових ЧП БШМС локальні та глобальні потокові характеристики можна визначати не лише для вузлів окремих шарів, але й для сукупності міжшарових взаємодій загалом. Виходячи з цього, параметри

$$\varsigma_{ij}^{in}(t) = \sum_{m=1}^{M} V_{ji}^{mm}(t), \quad \varsigma_{ij}^{out}(t) = \sum_{m=1}^{M} V_{ij}^{mm}(t), \qquad (5)$$

за аналогією з відповідними поняттями теорії зважених мереж [18] визначають вхідну та вихідну потокову силу взаємозв'язку між вузлами $n_i$ та $n_j$ загальної сукупності вузлів $\mathbf{V}^M$, враховуючи усі способи реалізації цього взаємозв'язку у різних шарах БШМС. Тоді параметри

$$\varsigma_{i}^{in}(t) = \sum_{j=1}^{N^M} \varsigma_{ji}^{in}(t) = \sum_{m=1}^{M} \delta_{m,i}^{in,intra}(t), \quad \varsigma_{i}^{out}(t) = \sum_{j=1}^{N^M} \varsigma_{ji}^{out}(t) = \sum_{m=1}^{M} \delta_{m,i}^{out,intra}(t), \quad i,j = \overline{1, N^M}, \quad t \geq T, \quad (6)$$

визначають вхідну та вихідну потокову агрегат-силу взаємозв'язків вузла $n_i$ загальної сукупності вузлів БШМС.

Оскільки ми розглядаємо монопотокові БШМС у яких міжшарові взаємодії можливі лише між вузлами з однаковими номерами у загальній сукупності вузлів, то матриці $\mathbf{V}^{mk}(t)$ мають діагональну структуру та діагональні елементи цих матриць $V_{ii}^{mk}(t)$ можуть бути відмінними від 0 тоді і тільки тоді, коли вузол $n_i$ входить до складу $m$-го та $k$-го шарів БШМС, $m \neq k$. Тоді потоковий агрегат-ступінь $\delta_i^{inter}(t)$ міжшарових взаємозв'язків вузла $n_i$ у багатошаровій мережевій системі визначається за співвідношенням

$$\delta_i^{in,inter}(t) = \sum_{m=1}^{M} \delta_{m,i}^{in,inter}(t), \quad \delta_i^{out,inter}(t) = \sum_{m=1}^{M} \delta_{m,i}^{out,inter}(t), \qquad (7)$$

у якому параметри

$$\delta_{m,i}^{in,inter}(t) = \sum_{k=1}^{M} V_{ii}^{mk}(t), \quad \delta_{m,i}^{out,inter}(t) = \sum_{k=1}^{M} V_{ii}^{km}(t), \quad t \geq T, \qquad (8)$$

визначають вхідні та вихідні міжшарові потокові ступені вузла $n_i$ у $m$-му шарі БШМС відповідно. Тоді загальні вхідний та вихідний потокові ступені вузла $n_i$ у процесі внутрішньо та міжсистемних взаємодій визначаються за співвідношеннями

$$\psi_i^{in}(t) = \varsigma_i^{in}(t) + \delta_i^{in,inter}(t), \quad \psi_i^{out}(t) = \varsigma_i^{out}(t) + \delta_i^{out,inter}(t), \quad t \geq T, \qquad (9)$$

та є функціональними узагальненнями поняття центральності за ступенем, яке визначається для вузлів багатошарової мережі [2]. Значення цих параметрів дозволяють у кількісному вимірі визначати, яким чином ураження вузла $n_i$ вплине на процес функціонування усіх суміжних із ним вузлів БШМС, $i = \overline{1, N^M}$.

Глобальними називатимемо характеристики елемента, які описують той або інший аспект його взаємодії з усіма іншими елементами цієї системи. Введемо у якості глобальних потокових характеристик вузлів МП ЧП БШМС параметри їх впливу на інші вузли багато-шарової системи. Нехай $v_k^{out}(t, n_i^m, n_j^l)$ – об'єм потоків, згенерованих у вузлі $n_i^m$ $m$-го шару та прийнятих у вузлі $n_j^l$ $l$-го шару БШМС, які пройшли шляхом $p_k(n_i^m, n_j^l)$ за період $[t-T, t]$, $t \geq T$, $K_{ij}^{ml}$ – кількість усіх можливих шляхів, які поєднують вузли $n_i^m$ та $n_j^l$, $k = \overline{1, K_{ij}^{ml}}$, $i, j = \overline{1, N^M}$, $m, l = \overline{1, M}$. Позначимо через

$$V^{out}(t, n_i^m, n_j^l) = \sum_{k=1}^{K_{ij}^{ml}} v_k^{out}(t, n_i^m, n_j^l) \tag{10}$$

сумарний об'єм потоків, згенерованих у вузлі $n_i^m$ та спрямованих для кінцевого прийняття у вузол $n_j^l$ БШМС усіма можливими шляхами за період $[t-T, t], t \geq T$. Параметр $V^{out}(t, n_i^m, n_j^l)$ визначає реальну силу впливу вузла $n_i^m$ на вузол $n_j^l$ на підставі сумарних об'ємів потоків, які надійшли з вузла $n_i^m$ у вузол $n_j^l$ за період тривалістю $T$, $i, j = \overline{1, N^M}$, $m, l = \overline{1, M}$.

Нехай $R_i^{m,l,out}(t) = \{j_{i_1}^l, ..., j_{i_{L_i^{ml}(t)}}^l\}$ – множина номерів вузлів $l$-го шару БШМС, які є кінцевими приймачами потоків, згенерованих у вузлі $n_i^m$, $L_i^{ml}(t)$ – кількість елементів множини $R_i^{m,l,out}(t)$, які також можуть змінюватися протягом періоду $[t-T, t], t \geq T$. Параметр

$$\xi_i^{m,l,out}(t) = \sum_{j \in R_i^{m,l,out}(t)} V^{out}(t, n_i^m, n_j^l) / s(\mathbf{V}^M(t)), \quad \xi_i^{m,l,out}(t) \in [0, 1], \tag{11}$$

визначає силу впливу вузла $n_i^m$ на $l$-тий шар-систему загалом, $t \geq T$, $i = \overline{1, N^M}$, $m, l = \overline{1, M}$. У формулі (11) значення

$$s(\mathbf{V}^M(t)) = \sum_{m,k=1}^{M} \sum_{i,j=1}^{N^M} V_{ij}^{mk}(t), \tag{12}$$

як сума усіх елементів матриці $\mathbf{V}^M(t)$, є глобальною потоковою характеристикою БШМС, рівною сумарним об'ємам потоків, які пройшли багатошаровою системою за період тривалістю $[t-T, t], t \geq T$. Потужність впливу вузла $n_i^m$ на $l$-тий шар-систему визначимо за допомогою параметра

$$p_i^{m,l,out}(t) = L_i^{ml}(t) / N^M, \quad p_i^{m,l,out} \in [0, 1],$$

а множину $R_i^{m,l,out}(t)$ називатимемо областю впливу вузла $n_i^m$ на *l*-тий шар-систему БШМС. Параметри $\xi_i^{m,l,out}(t)$, $p_i^{m,l,out}(t)$ та $R_i^{m,l,out}(t)$ називатимемо вихідними параметрами впливу вузла $n_i^m$ на *l*-тий шар-систему БШМС, $i = \overline{1, N^M}$, $m, l = \overline{1, M}$.

Аналогічно до вихідних визначаються параметри $\xi_i^{l,m,in}(t)$, $p_i^{l,m,in}(t)$ та $R_i^{l,m,in}(t)$ сили, потужності та області вхідного впливу, які називатимемо вхідними параметрами впливу *l*-го шару на вузол $n_i^m$ багатошарової мережевої системи, $[t-T, t], t \geq T$. Значення параметрів вхідного та вихідного впливу вузла $n_i^m$ на *l*-тий шар дозволяють у кількісному вимірі визначати, яким чином ураження цього вузла вплине на процес функціонування на *l*-го шару БШМС, $i = \overline{1, N^M}$, $m, l = \overline{1, M}$.

Вихідну силу впливу вузла $n_i^m$ на всю багатошарову мережеву систему протягом періоду $[t-T, t], t \geq T$, обчислюємо за формулою

$$\xi_i^{m,out}(t) = \sum_{l=1}^{M} \xi_i^{m,l,out}(t)/M, \quad \xi_i^{m,out}(t) \in [0,1], \qquad (13)$$

у якій значення $\xi_i^{m,l,out}(t)$ визначається за формулою (11). Область вихідного впливу $R_i^{m,out}(t)$ вузла $n_i^m$ на БШМС визначається за співвідношенням

$$R_i^{m,out}(t) = \bigcup_{l=1}^{M} R_i^{m,l,out}(t).$$

Тоді потужність $p_i^{m,out}(t)$ вихідного впливу вузла $n_i^m$ на БШМС дорівнює відношенню кількості елементів множини $R_i^{m,out}(t)$ до значення $N^M$. Аналогічно до вихідних визначаються сила $\xi_i^{m,in}(t)$, область $R_i^{m,in}(t)$ та потужність $p_i^{m,in}(t)$ вхідного впливу БШМС на вузол $n_i^m$ протягом періоду часу $[t-T, t], t \geq T$, $i = \overline{1, N^M}$, $m = \overline{1, M}$. Зазначимо, що параметри вхідного та вихідного впливу, зокрема області впливу вузла, на відміну від його глобальних структурних характеристик, дають змогу визначати участь цього вузла у процесі міжсистемних взаємодій навіть якщо він входить до складу лише одного шару, тобто не є точкою переходу БШМС. Ураження вузла-генератора потоків означає необхідність пошуку для кінцевих приймачів нового джерела постачання, а вузла-приймача – пошуку для виробників нових ринків збуту, що призведе до принаймні тимчасових труднощів у їх роботі. Параметри впливу дають змогу визначати, до яких втрат у кількісному вимірі (за вартістю потоків, рух яких не був здійснений мережею) це призведе та на скільки елементів внутрішньо та міжсистемних взаємодій пошириться [19, 20].

Вихідну силу впливу *m*-го шару на *l*-тий шар БШМС протягом періоду $[t-T, t], t \geq T$, обчислюємо за формулою

$$\xi^{m,l,out}(t) = \sum_{i=1}^{N^M} \xi_i^{m,l,out}(t) / N^M, \quad \xi^{m,l,out}(t) \in [0,1], \qquad (14)$$

у якій значення $\xi^{m,l,out}(t)$ обчислюється згідно (11). Область $R^{m,l,out}(t)$ вихідного впливу *m*-го шару на *l*-тий шар багатошарової системи визначається за співвідношенням

$$R^{m,l,out}(t) = \bigcup_{i=1}^{N^M} R_i^{m,l,out}(t),$$

а потужність цього впливу $p^{m,l,out}(t)$ дорівнює відношенню кількості елементів області $R^{m,l,out}(t)$ до значення $N^M$. Аналогічно визначаються вхідна сила $\xi^{m,l,in}(t)$, область $R^{m,l,in}(t)$ та потужність $p^{m,l,in}(t)$ впливу *m*-го шару на *l*-тий шар БШМС протягом періоду, $m \neq l$, $m, l = \overline{1, M}$, $[t-T, t], t \geq T$.

Вихідну силу впливу *m*-го шару на БШМС загалом обчислюємо за формулою

$$\xi^{m,out}(t) = \sum_{\substack{l=1 \\ l \neq m}}^{M} \xi^{m,l,out}(t) / (M-1), \quad \xi^{m,out}(t) \in [0,1], \qquad (15)$$

у якій значення $\xi^{m,l,out}(t)$ обчислюється згідно (14), а область цього впливу визначається за співвідношенням

$$R^{m,out}(t) = \bigcup_{\substack{l=1 \\ l \neq m}}^{M} R^{m,l,out}(t).$$

Потужність вихідного впливу *m*-го шару на БШМС загалом дорівнює відношенню кількості елементів області $R^{m,out}(t)$ до значення $N^M$. Аналогічно визначаються сила $\xi^{m,in}(t)$, область $R^{m,in}(t)$ та потужність $p^{m,in}(t)$ вхідного впливу *m*-го шару на БШМС загалом протягом періоду $[t-T, t], t \geq T$. Значення параметрів вхідного та вихідного впливу *m*-го шару на *l*-тий шар або БШМС загалом дозволяють у кількісному вимірі визначати, яким чином ураження цього шару вплине на процес функціонування на *l*-го шару та багатошарової мережевої системи в цілому, $m, l = \overline{1, M}$.

Наступним типом глобальних потокових характеристик елементів БШМС є їх параметри посередництва. Ці характеристики визначають важливість вузла або ребра багатошарової мережевої системи у забезпеченні руху транзитних потоків під час внутрішньо та міжсистем-них взаємодій [4]. Для скорочення викладу зосередимося на визначенні параметрів посередництва точок переходу БШМС, як найважливіших елементів, які забезпечують міжшарові взаємодії у монопотокових частково покритих багатошарових системах. Враховуючи, що ми розглядаємо випадок, коли міжшарові зв'язки можливі лише між вузлами з однаковими номерами із загальної сукупності $V^M$ вузлів БШМС, позначимо через $P_{i,ml}^{K_i^{ml}} = \{p_{i,ml}^k\}_{k=1}^{K_i^{ml}}$ – сукупність шляхів, що поєднують вузли-генератори та вузли-приймачі потоків $l$-го та $m$-го шарів БШМС, та проходять через точку переходу $n_i^{ml}$, яка входить до складу обох цих шарів. Нехай $v_{i,ml}^k(t)$ – об'єм потоків, які пройшли шляхом $p_{i,ml}^k$ від вузла-генератора до вузла-приймача, а отже і через точку переходу $n_i^{ml}$, за період $[t-T,t], t \geq T$, $i = \overline{1,N^M}$, $m,l = \overline{1,M}$. Тоді величина

$$V_{i,ml}^{K_i^{ml}}(t) = \sum_{k=1}^{K_i^{ml}} v_{i,ml}^k(t)$$

визначає сумарний об'єм потоків, які пройшли сукупністю шляхів $P_{i,ml}^{K_i^{ml}}$, а отже і через точку переходу $n_i^{ml}$ за цей же проміжок часу. Величину

$$\Phi_i^{ml}(t) = V_{i,ml}^{K_i^{ml}}(t) / s(\mathbf{V}^M(t)), \Phi_i^{ml}(t) \in [0,1], \quad (16)$$

яка визначає питому вагу в системі потоків, що проходять через точку переходу $n_i^{ml}$ за період $[t-T,t], t \geq T$, називатимемо мірою посередництва цієї точки переходу в процесі взаємодії $l$-го та $m$-го шарів БШМС. Множину $M_i^{ml}$ усіх вузлів $l$-го та $m$-го шарів БШМС, які є генераторами та кінцевими приймачами потоків, що проходять транзитом через вузол $n_i^{ml}$ шляхами із сукупності $P_{i,ml}^{K_i^{ml}}$, називатимемо областю посередництва, а відношення $\eta_i^{ml}$ кількості вузлів множини $M_i^{ml}$ до значення $N^M$ – потужністю посередництва точки переходу $n_i^{ml}$, $i = \overline{1,N^M}$, $m,l = \overline{1,M}$.

Параметри посередництва точки переходу $n_i^m$ у процесі міжсистемних взаємодій у межах всієї БШМС визначимо наступним чином. Міру посередництва $\Phi_i^m(t)$ точки переходу $n_i^m$ у всій багатошаровій системі обчислимо за формулою

$$\Phi_i^m(t) = \sum_{\substack{l=1 \\ l \neq m}}^{M} \Phi_i^{ml}(t)/(M-1), \Phi_i^m(t) \in [0,1], \quad (17)$$

у якій значення $\Phi_i^{ml}(t)$ обчислюється згідно (16). Область посередництва точки переходу $n_i^m$ у всій БШМС визначається за співвідношенням

$$M_i^m(t) = \bigcup_{\substack{l=1 \\ l \neq m}}^{M} M_i^{ml}(t).$$

Тоді потужність посередництва $N_i^m(t)$ точки переходу $n_i^m$ у всій БШМС дорівнює відношенню кількості елементів множини $M_i^m(t)$ до значення $N^M$. Зазначимо, що для вузлів, які не є точками переходу БШМС, параметри міри, області та потужності посередництва визначаються за такими ж принципами. Аналогічно можна визначити параметри міри $\Phi_{ij}^m(t)$, області $M_{ij}^m(t)$ та потужності $N_{ij}^m(t)$ посередництва ребра $(n_i^m, n_j^m)$ $m$-го шару БШМС, $i,j = \overline{1, N^M}$, $m = \overline{1, M}$, $[t-T, t], t \geq T$. Це, зокрема, означає, що, на відміну від структурних характеристик, параметри посередництва елементів БШМС дають змогу встановлювати участь у міжсистемних взаємодіях навіть тих вузлів та ребер, які входять до складу лише одного шару багатошарової мережевої системи. Значення параметрів посередництва вузла $n_i^m$ у БШМС дозволяють у кількісному вимірі визначати, яким чином ураження цього вузла вплине на процес міжсистемних взаємодій загалом, $i = \overline{1, N^M}$, $m = \overline{1, M}$.

Міру посередництва $m$–го шару в межах всієї БШМС протягом періоду $[t-T, t], t \geq T$ обчислюємо за формулою

$$\Phi^m(t) = \sum_{i=1}^{N^M} \Phi_i^m(t) / N^M, \Phi^m(t) \in [0,1], \qquad (18)$$

у якій значення $\Phi^m(t)$ обчислюється згідно (17). Очевидно, що область посередництва $m$-го шару у БШМС загалом визначається із співвідношення

$$M^m(t) = \bigcup_{i=1}^{N^M} M_i^m(t),$$

а потужність посередництва $N^m(t)$ дорівнює відношенню кількості елементів області $M^m(t)$, $[t-T, t], t \geq T$, до значення $N^M$. Значення параметрів посередництва $m$–го шару у БШМС дозволяють у кількісному вимірі визначати, яким чином ураження цього шару вплине на процес міжсистемних взаємодій загалом, $m = \overline{1, M}$.

Важливість вузла $n_i$ загальної сукупності вузлів БШМС як генератора, кінцевого приймача або транзитера потоків обчислюємо за формулами

$$\xi_i^{out}(t) = \sum_{m=1}^{M} \xi_i^{m,out}(t)/M, \quad \xi_i^{out}(t) \in [0,1], \tag{19}$$

$$\xi_i^{in}(t) = \sum_{m=1}^{M} \xi_i^{m,in}(t)/M, \quad \xi_i^{in}(t) \in [0,1], \tag{20}$$

$$\Phi_i(t) = \sum_{m=1}^{M} \Phi_i^m(t)/M, \Phi_i(t) \in [0,1], i = \overline{1, N^M}, [t-T, t], t \geq T, \tag{21}$$

відповідно. Області вхідного $R_i^{in}(t)$, вихідного $R_i^{out}(t)$ впливу та посередництва $M_i(t)$ вузла $n_i$ у БШМС визначатимемо за співвідношеннями

$$R_i^{in}(t) = \bigcup_{m=1}^{M} R_i^{m,in}(t), \, R_i^{out}(t) = \bigcup_{m=1}^{M} R_i^{m,out}(t), \, M_i(t) = \bigcup_{m=1}^{M} M_i^m(t),$$

а потужності вхідного $p_i^{in}(t)$, вихідного $p_i^{out}(t)$ впливу та посередництва $N_i(t)$ вузла $n_i$ на БШМС загалом, як відношення кількості елементів областей $R_i^{in}(t)$, $R_i^{out}(t)$ та $M_i(t)$, $i = \overline{1, N^M}$, до значення $N^M$ відповідно.

На підставі параметрів вхідного та вихідного впливу, а також посередництва вузла $n_i$ загальної сукупності вузлів $\mathbf{V}^M$ ми можемо визначити глобальний показник взаємодії цього вузла із БШСМ загалом, а саме, параметр $\Xi_i(t)$ сили взаємодії вузла $n_i$ із багатошаровою системою, який обчислюємо за формулою

$$\Xi_i(t) = (\xi_i^{out}(t) + \xi_i^{in}(t) + \Phi_i(t))/3, \quad t \geq T, \tag{22}$$

та визначає його сукупну роль у багатошаровій системі як генератора, кінцевого приймача та транзитера потоків; область $\Omega_i(t)$ взаємодії вузла $n_i$ з БШМС, яка визначається за співвідношенням

$$\Theta_i(t) = R_i^{in}(t) \bigcup R_i^{out}(t) \bigcup M_i(t)$$

та потужність взаємодії вузла $n_i$ із БШМС, яка дорівнює відношенню кількості елементів області $\Omega_i(t)$, $t \geq T$, до значення $N^M$. Зрозуміло, що параметри взаємодії вузла $n_i$, $i = \overline{1, N^M}$, із БШМС визначають рівень їх залежності один від одного та дають змогу встановлювати, яким чином ураження цього вузла вплине на процес міжсистемних взаємодій загалом.

Важливість *m*-го шару як генератора, кінцевого приймача або транзитера потоків у багатошаровій мережевій системі обчислюємо за формулами

$$\xi^{m,out}(t) = \sum_{i=1}^{N^M} \xi_i^{m,out}(t)/N^M, \quad \xi^{m,out}(t) \in [0,1], \tag{23}$$

$$\xi^{m,in}(t) = \sum_{i=1}^{N^M} \xi_i^{m,in}(t)/N^M, \quad \xi^{m,in}(t) \in [0,1], \qquad (24)$$

$$\Phi^m(t) = \sum_{i=1}^{N^M} \Phi_i^m(t)/N^M, \quad \Phi^m(t) \in [0,1], \qquad (25)$$

відповідно. Області вхідного $R^{m,in}(t)$, вихідного $R^{m,out}(t)$ впливу та посередництва $M^m(t)$ $m$-го шару у БШМС визначатимемо за співвідношеннями

$$R^{m,in}(t) = \bigcup_{i=1}^{N^M} R_i^{m,in}(t), \; R^{m,out}(t) = \bigcup_{i=1}^{N^M} R_i^{m,out}(t), \; M^m(t) = \bigcup_{i=1}^{N^M} M_i^m(t),$$

а потужності вхідного $p^{m,in}(t)$, вихідного $p^{m,out}(t)$ впливу та посередництва $N^m(t)$ $m$-го шару у БШМС загалом, як відношення кількості елементів областей $R^{m,in}(t)$, $R^{m,out}(t)$ та $M^m(t)$, $m = \overline{1,M}$, $[t-T,t], t \geq T$, до значення $N^M$ відповідно.

На підставі параметрів вхідного та вихідного впливу, а також посередництва $m$-го шару ми можемо сформувати глобальний показник взаємодії цього шару із БШСМ загалом, а саме, параметр $\Xi^m(t)$ сили взаємодії $m$-го шару із багатошаровою системою, який обчислюється за формулою

$$\Xi^m(t) = (\xi^{m,out}(t) + \xi^{m,in}(t) + \Phi^m(t))/3, \quad t \geq T, \qquad (26)$$

та визначає його сукупну роль у багатошаровій системі як генератора, кінцевого приймача та транзитера потоків; область $\Omega^m(t)$ взаємодії $m$-го шару з БШМС, яка визначається за співвідношенням

$$\Theta^m(t) = R^{m,out}(t) \bigcup R^{m,out}(t) \bigcup M^m(t)$$

та потужність взаємодії $m$-го шару із БШМС, яка дорівнює відношенню кількості вузлів області $\Omega^m(t)$, $m = \overline{1,M}$, $t \geq T$, до значення $N^M$. Зрозуміло, що параметри взаємодії шару із багатошаровою мережевою системою визначають важливість цього шару у процесі міжсистемних взаємодій та дають змогу встановлювати, яким чином ураження цього шару вплине на процес функціонування БШМС.

Загалом під потоковою моделлю багатошарової мережевої системи ми розуміємо не тільки її матрицю суміжності виду (2), але й усю сукупність локальних та глобальних потокових характеристик елементів та шарів БШМС, визначених у співвідношеннях (3)-(26). Це пояснюється тим, що за допомогою показників функціональної важливості можна визначити роль цих елементів та шарів у процесі внутрішньо та міжсистемних взаємодій, особливості їх поведінки залежно від процесів, які розгортаються у багатошарових системах, зокрема, на яку область БШМС пошириться ураження окремого вузла, ребра або шару та до яких втрат це призведе тощо. Прогнозування значень потокових характеристик дають змогу визначити принаймні короткострокові тенденції зміни реальної важливості елементів та шарів у процесі міжсистемних взаємодій, що може суттєво вплинути на пріоритети захисту БШМС. Слід також враховувати, що здійснення масштабної групової або загальносистемної атаки потребує значних зусиль не лише об'єкта для захисту та подоланню наслідків такої атаки, але й суб'єкта для мінімізації витрат на її реалізацію. Різні рівні (вузла на вузол, шар або БШМС загалом) та види (вхідні, вихідні, транзитні) взаємодії дозволяють оптимізувати такі витрати під час побудови сценаріїв цілеспрямованих атак на багатошарові системи. Так, великий автовиробник або виробник високоточної зброї є кінцевим приймачем великої кількості комплектуючих. При цьому для припинення виготовлення ними високотехнологічної продукції достатньо лише припинити постачання сучасних інтегральних мікросхем (чипів) [21]. Прибутки від продажу нафти та газу формують від 40 до 50 відсотків бюджету рф. Ембарго на їх постачання суттєво вплине на сировинний сектор, який складає до 40% всієї економіки цієї країни [22]. Розрив логістичних ланцюжків постачання боєприпасів суттєво впливає на інтенсивність наступальних дій ворога тощо. Тобто, для блокування або суттєвого ускладнення роботи найважливіших складових системи достатньо усунути лише одне ребро постачання з відповідного вузла-генератора або знищити вузол-транзитер потоків. Звідси також слідує, що диверсифікація як джерел постачання, так і кінцевих приймачів, а також достатній резерв альтернативних шляхів руху потоків є дієвими засобами захисту БШМС від цілеспрямованих атак різних типів.

Як і для одношарових МС, використання потокової моделі БШМС дозволяє значно ефективніше порівняно зі структурною вирішувати цілу низку практично важливих проблем системного аналізу, а саме [23]

1. Виділяти найважливіші з функціонального погляду складові БШМС, що суттєво сприяє кращому розумінню процесів, які перебігають у них, і унаслідок цього подоланню проблеми складності системних досліджень, оскільки зосереджує першочергову увагу саме на визначальних об'єктах внутрішньо та міжсистемних взаємодій.

2. Здійснювати пошук фіктивних та прихованих елементів БШМС. Видалення зі складу структури вузлів та ребер багатошарової системи, які практично не приймають участі у процесі її функціонування дозволяє суттєво зменшувати розмірність потокової моделі, а включення прихованих елементів – значно підвищувати її адекватність.

3. Досліджувати процеси, які розгортаються у БШМС не лише у період росту [24], але й на усіх етапах життєвого циклу як окремих шарів-систем, так і багатошарової системи загалом. При цьому потоковий підхід дає змогу визначати не лише напрямки розвитку або згортання структури та процесу функціонування окремих складових БШМС, але й виявляти причини таких явищ (впровадження новітніх технологій призводить до зникнення багатьох застарілих промислових систем, систем постачання та зв'язку тощо).

4. Визначати альтернативні шляхи руху через БШМС в умовах критичного завантаження потоками окремих шарів-систем або ураження цих шарів.

5. Здійснювати пошук спільнот у БШМС, особливо у випадках, коли вони також є багатошаровими утвореннями та слабо виявляються в окремих шарах-системах.

6. Розробляти засоби захисту від цілеспрямованих атак або протидії поширенню нецільових уражень як внутрішньо, так і міжсистемних взаємодій та кількісно визначати наслідки таких уражень.

**ВИСНОВКИ**

У статті запропоновано підхід до декомпозиції багатовимірних частково покритих мережевих систем на монопотокові БШМС та побудована потокова модель таких систем, яка описує процеси внутрішньо та міжсистемних взаємодій у надсистемних утвореннях, які забезпечують рух потоків певного типу. Визначені основні локальні та глобальні потокові характеристики вузлів, ребер та окремих шарів БШМС, які дають змогу встановлювати їх важливість у процесі функціонування багатошарових мережевих систем. Показані можливості застосування пропонованої моделі для вирішення низки практично важливих задач системного аналізу. Наступним етапом наших досліджень буде побудова ефективних сценаріїв послідовних і одночасних групових та загальносистемних атак на монопотокові частково покриті багатошарові мережеві системи на підставі потокового підходу до моделювання процесу міжсистемних взаємодій та визначення функціональних показників важливості його складових.